\documentclass[prb,twocolumn, superscriptaddress, amsmath, amssymb,reprint,nobibnotes,letterpaper]{revtex4-1}
\usepackage{graphicx}% Include figure files
\usepackage{dcolumn}% Align table columns on decimal point
\usepackage{bm}% bold math
\usepackage{upgreek}% Greak symbols NOT italic
\usepackage{times}
\usepackage{mathrsfs}
\usepackage{multirow}
\usepackage{fancyhdr} % header/footers
\usepackage{hyperref} % hyperlinks
\usepackage{mathtools}
\usepackage[version=3]{mhchem}

\begin{document}

%\preprint{ABC}
\title{One-dimensional electron gas in strained lateral heterostructures of single layer materials}
\author{O.~Rubel}
\email[e-mail:~]{rubelo@mcmaster.ca}
\affiliation{Department of Materials Science and Engineering, McMaster University, 1280 Main Street West,
Hamilton, Ontario L8S 4L8, Canada}

\date{\today}

\begin{abstract}
\noindent Confinement of the electron gas along one of the spatial directions opens an avenue for studying fundamentals of quantum transport along the side of numerous practical electronic applications, with high-electron-mobility transistors being a prominent example. A heterojunction of two materials with dissimilar electronic polarisation can be used for engineering of the conducting channel. Extension of this concept to single-layer materials leads to one-dimensional electron gas (1DEG). \ce{MoS2}/\ce{WS2} lateral heterostructure is used as a prototype for the realisation of 1DEG. The electronic polarisation discontinuity is achieved by straining the heterojunction taking advantage of dissimilarities in the piezoelectric coupling between \ce{MoS2} and \ce{WS2}. A complete theory that describes an induced electric field profile in lateral heterojunctions of two-dimensional materials is proposed and verified by first principle calculations. 
\end{abstract}

\pacs{TBD}

\maketitle

%-----------------------------------------------------------------------
%
%                       I N T R O D U C T I O N
%
%-----------------------------------------------------------------------
%\section{Introduction}\label{Sec:Introduction}

\noindent Confinement of electrons along one of the spatial directions results in a two-dimensional electron gas (2DEG) that exhibits interesting physical phenomena along the side of useful technological applications. Particular examples include the field of quantum
transport and mesoscopic physics \cite{Klitzing_PRL_45_1980} as well as high-electron-mobility transistors that are used in integrated circuits as digital on-off switches \cite{Dimitrijev_MB_40_2015}. The advantage of 2DEG conducting channel is the high mobility of charge carriers due to the absence of deleterious effects inherent to ionised impurity scattering that allows for ballistic transport \cite{Kumar_PRL_105_2010}. Engineering of 2DEG conventionally requires the use of a modulation doping technique \cite{Dingle_APL_33_1978} as in the case of (AlGa)As/GaAs heterostructures. Alternatively, the 2DEG can be achieved in undoped structures with an extreme band bending induced by the strong electric field at a heterojunction between two dielectric materials with dissimilar electronic polarisation such as (AlGa)N/GaN interface \cite{Khan_APL_60_1992,Ambacher_JAP_87_2000}. It is interesting to see whether polarisation effects in two-dimensional (2D) materials can be used to achieve confinements of electrons along one spatial direction?

2D materials become a perspective avenue for keeping up with latest trends in miniaturisation of electronics, culminating in a demonstration of the single layer \ce{MoS2} transistor \cite{Radisavljevic_NN_6_2011,Kim_NC_3_2012,Radisavljevic_NM_12_2013}. Unlike group III-nitrides, free-standing transition-metal dichalcogenides do not possess spontaneous polarisation due to symmetry arguments. However, single-atomic-layer h-BN and monolayer transition-metal dichalcogenides have been theoretically predicted \cite{Duerloo_JPCL_3_2012} and experimental confirmed \cite{Wu_N_514_2014,Zhu_NN_10_2015} to show piezoelectricity as a result of strain-induced lattice distortions. Two types of heterostructures that involve 2D materials are discussed in the literature: (i) multilayer heterostructures produced by stacking of different 2D materials, so-called van der Waals heterostructures \cite{Geim_N_499_2013}, and (ii) lateral heterostructures, which are formed when two materials are covalently bonded \textit{within} the 2D plane \cite{Huang_NM_13_2014}.

It will be shown that a lateral heterojunction of 2D materials with dissimilar piezoelectric properties can be used to achieve additional confinement of charge carriers along the interface, which creates conditions for realisation of a one-dimensional electron gas (1DEG). A complete theory that describes an induced electric field profile in lateral heterojunctions of 2D materials is presented and verified by first principle calculations.

%-----------------------------------------------------------------------
%
%                       R E S U L T S
%
%-----------------------------------------------------------------------
%\section{Results and discussion}\label{Sec:Results}

\subsection*{\large First-principle model \hfill~}\vspace{-12pt}

\begin{figure*}
    \includegraphics[width=0.8\textwidth]{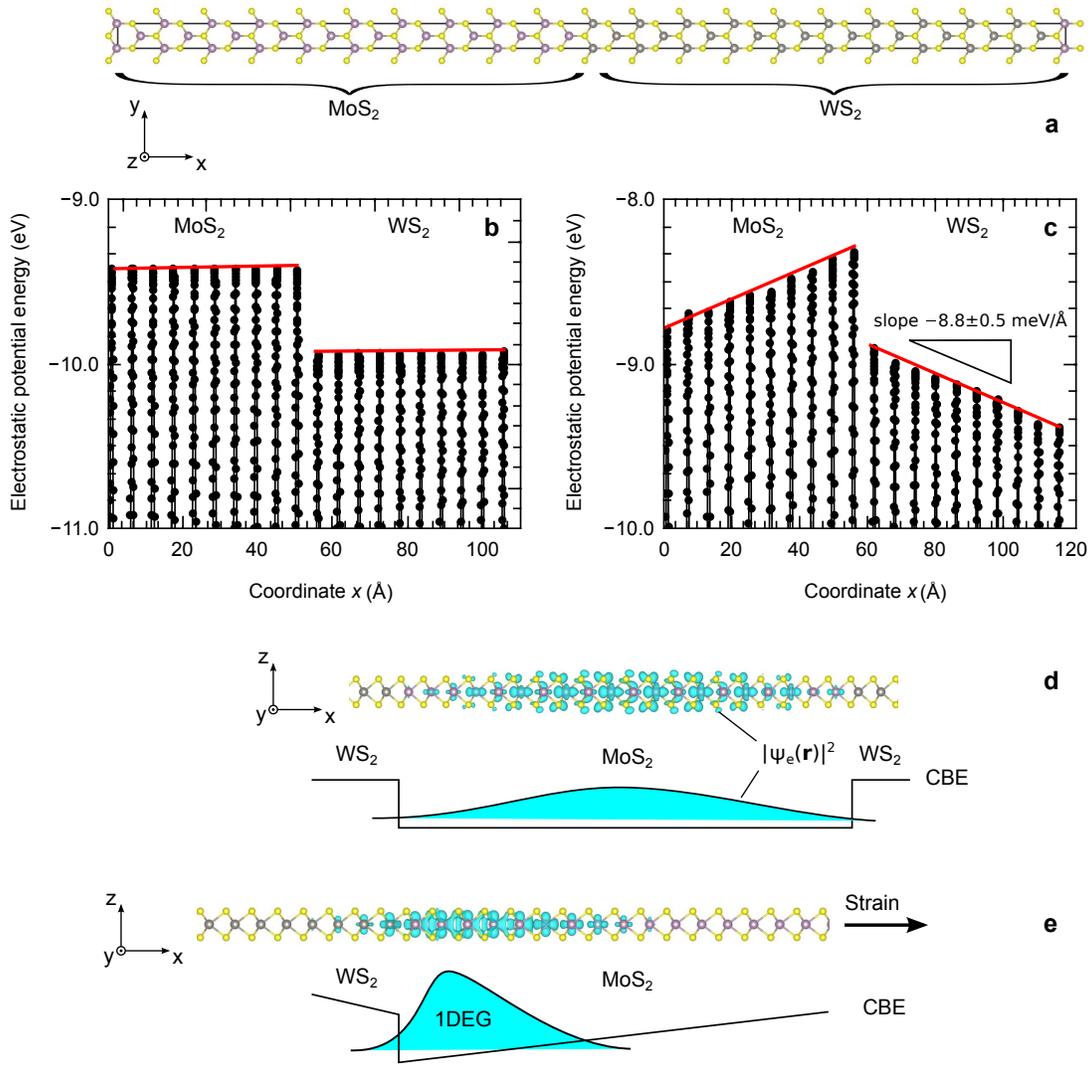}\\
	\caption{Electron confinement in lateral \ce{MoS2}/\ce{WS2} heterojunction. (a) 80-atoms model of the heterojunction. (b,c) Electrostatic potential energy profile across the heterojunction without strain and with the strain of $\epsilon_{1}=0.1$, respectively. The scan is taken between points with the fractional coordinates $(0,1/2,0)$ and $(1,1/2,0)$. The built-in electric field corresponds to a macroscopic slope of the potential energy.  (d,e) The electron wavefunction amplitude $|\psi_\text{e}(\bm{r})|^2$ represents the lowest unoccupied state in unstrained and strained heterostructures, respectively. The strain-induced electric field confines electrons forming a one-dimensional conducting channel along the \ce{MoS2}/\ce{WS2} interface. The band diagrams show the spatial evolution of the conduction band edge (CBE) schematically to assist with interpretation of the wavefunction plot.}\label{Fig:Energy-profile}
\end{figure*}

\noindent First, we will use an \textit{ab initio} model to explore the feasibility of achieving conditions for 1D confinement of charge carriers in a lateral heterojunction of two single-layer materials. For this purpose, an 80-atoms \ce{MoS2}/\ce{WS2} supercell is constructed as illustrated in Fig.~\ref{Fig:Energy-profile}(a). \ce{MoS2} and \ce{WS2} are chosen due to an almost identical lattice parameter of two materials (less that 0.1\% mismatch), which reduces the misfit strain at the interface. One would expect the heterostructure to possess no built-in electric field since transition metal dichalcogenides manifest no net polarisation unlike group-III nitride bulk semiconductors. This hypothesis can be verified by plotting the potential energy across the heterojunction (Fig.~\ref{Fig:Energy-profile},~b). The potential energy profile shows periodic oscillations with minima in the vicinity of nuclei and maxima corresponding to interstitial regions. It is evident that maxima of the potential energy remain constant within \ce{MoS2} and \ce{WS2} domains with an abrupt step-like transition at the interface. The confinement of charge carriers resembles that in a quantum well (Fig.~\ref{Fig:Energy-profile},d).

Next, the same heterostructure is uniformly strained in the direction perpendicular to the heterojunction, i.e., along $x$-axis (Fig.~\ref{Fig:Energy-profile},~a). The magnitude of strain is deliberately chosen high (10\%) in order to magnify observed effects. The Poisson's contraction is simulated by relaxing the second lateral dimension of the supercell to eliminate the macroscopic stress $\sigma_{22}$, accompanied by a full relaxation of internal degrees of freedom.   It is found that, after relaxation, the macroscopic strain of 10\% is non-uniformly distributed among both material domains. The effective strain in \ce{MoS2} is 10.5\%, while \ce{WS2} accommodates only 9.5\%. This result can be attributed to differences in stiffness between two materials.

\begin{figure}
    \includegraphics[width=0.48\textwidth]{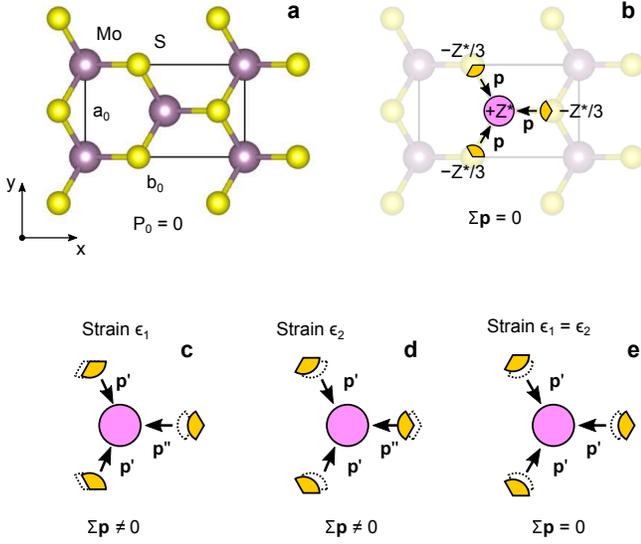}\\
	\caption{Strain-induced change in electronic polarisation of hexagonal \ce{MoS2}. (a) Rectangular unit cell. (b) Cancellation of local dipoles $\bm{p}$ induced by of the charge transfer $\pm Z^*$ due to the C$_3$ rotational symmetry, which results in the vanishing of a spontaneous polarisation $P_0$. (c,d) Symmetry breaking due to uniaxial strain induces a macroscopic dipole moment giving rise to the strain-induced polarisation. (e) Under the equal plain strain condition ($\epsilon_{1}=\epsilon_{2}$), the symmetry is preserved. Thus no change in polarisation should be observed.}\label{Fig:Polarization}
\end{figure}

It is also noticed that the external strain breaks 3-fold rotational symmetry, which is responsible for the absence of spontaneous polarisation in \ce{MoS2} and \ce{WS2} due to the cancellation of polarisation dipoles (Fig.~\ref{Fig:Polarization}). The symmetry breaking is evident from the disparity in Mo-S bond lengths: 2.52~{\AA} \textit{vs} 2.41~{\AA} for the bonds oriented along or tilted with respect to the strain direction. The electrostatic potential profile plotted in Fig.~\ref{Fig:Energy-profile}(c) reveals the presence of an electric field in \ce{MoS2} and \ce{WS2} domains of approximately equal magnitude, but the opposite direction. The magnitude of electric field varies ($\pm10$\%) depending on the coordinates of the line scan (see Supplementary information for more details); the average field is approximately $8.2\pm0.5$~mV/{\AA}. The created saw-like potential confines charge carriers in the vicinity of the \ce{MoS2}/\ce{WS2} interface (Fig.~\ref{Fig:Energy-profile},e) producing a narrow 1D conduction channel along $y$-axis of the width a few interatomic spacings.

Qualitatively, an origin of the electric field can be attributed to heterogeneity in polarisation induced by the strain in \ce{MoS2} and \ce{WS2} domains (see Fig.~\ref{Fig:Heterostructure}). To gain a quantitative understanding of the observed effects in 2D materials, a model that couples continuum mechanics and Poisson equation is developed below.

\subsection*{\large Continuum model \hfill~}\vspace{-12pt}

\noindent The purpose of this model is to describe the electric field profile induced due to piezoelectric effects in 2D strained heterostructures. The problem is similar to that solved by Ambacher~\textit{et al.}~\cite{Ambacher_JAP_87_2000} for AlGaN/GaN heterostructures, however, there are peculiarities related to 2D character of the materials in question, which warrant repeating some basic steps.

\begin{figure}
    \includegraphics[width=0.48\textwidth]{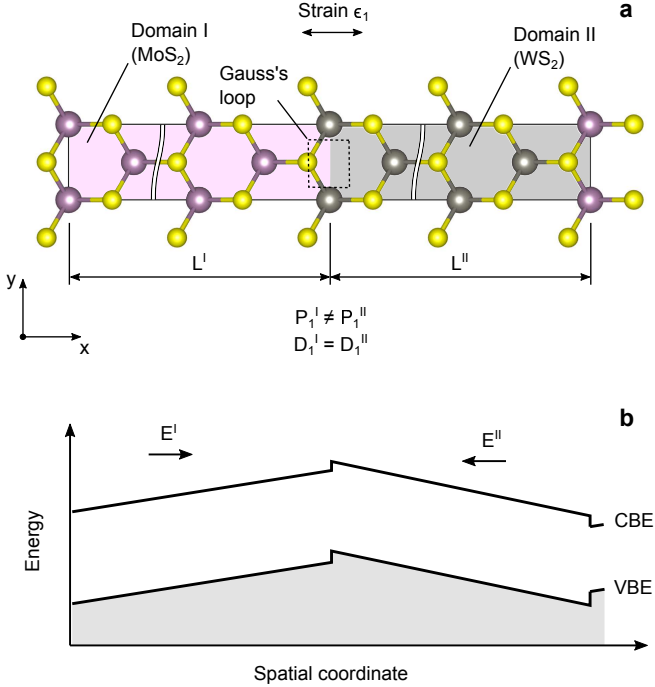}\\
	\caption{Continuum model of lateral \ce{MoS2}/\ce{WS2} heterointerface. (a) The strain along $x$-axis induces a heterogeneity of electronic polarisation $P$ due to differences in the piezoelectric response between two materials. (b) Inhomogeneities in polarisation create regions with an opposite electric field $E$ that results in subsequent spatial bending of the conduction band minima (CBM) and valence maxima (VBM).}\label{Fig:Heterostructure}
\end{figure}

The free electro-elastic energy density stored in a linear medium can be expressed as \cite{meitzler1988ieee}
\begin{equation}\label{Eq:w}
	w(\bm{\epsilon},\bm{E}) = 
	\frac{1}{2} \sum_i \sum_j C_{ij} \epsilon_i \epsilon_j + 
	\frac{1}{2} \sum_l \sum_m \varepsilon_{lm} E_lE_m,
\end{equation}
where $\bm{\epsilon}=(\epsilon_1,~\epsilon_2,~\epsilon_6)$ are components of the strain tensor written in the Voigt's matrix notations, $E_i$ is the electric field projection along $i$ axis, $C_{ij}$ are components of the stiffness matrix, $\varepsilon_{lm}$ are components of the electrical permittivity tensor of the material, and the range of indices $i,j=1,2,6$, $l,m=1,2$ is adapted to 2D. Oftentimes, the macroscopic strain is found by minimising the elastic energy only \cite{Ambacher_JAP_87_2000} (first term in Eq.~(\ref{Eq:w})). However, it should be emphasised that the electric field and strain are coupled through the electric displacement, which takes the form 
\begin{equation}\label{Eq:D_l}
	D_l = P_{0,l} + \sum_i e_{li}\epsilon_{i}  + \sum_m \varepsilon_{lm} E_m~.
\end{equation}
Here $\bm{P}_{0}$ is the permanent (spontaneous) polarisation and $e_{li}$ are components of piezoelectric strain tensor. In the absence of free charges, the Gauss's law requires
\begin{equation}\label{Eq:Gauss's_law}
	\nabla \cdot \bm{D} = 0.
\end{equation}
This implies continuity of the electric displacement at the interface of two domains (see Fig.~\ref{Fig:Heterostructure},a)
\begin{equation}\label{Eq:D_continuity}
	\bm{D}^\text{I} = \bm{D}^\text{II},
\end{equation}
which includes contributions from permanent, strain-induced, and field-induced electric dipoles in the material. The strain $\bm{\epsilon}(\bm{r})$ and electric field $\bm{E}(\bm{r})$ distributions can be found by minimising the total electro-elastic energy
\begin{equation}\label{Eq:W}
	W = \int w(\bm{r})~d\bm{r},
\end{equation}
subject to boundary conditions, e.g., an applied macroscopic strain.

2D materials pose a challenge related to defining the integration volume required to evaluate the total free energy in Eq.~(\ref{Eq:W}). There are attempts in the literature \cite{Huang_PRB_74_2006} to assign an effective thickness to atomically thin monolayers to compare their properties (strength, elastic modulus, or dielectric constant) to bulk materials. However, such analysis always bares the element of ambiguity. Alternatively, it seems more logical for 2D materials to use area rather than volume for normalising their specific properties. As a result, the stiffness coefficients $C$ acquire units of N/m, whereas the piezoelectric coefficients $e$ are expressed in units of C/m in 2D \cite{Duerloo_JPCL_3_2012}. To remain consistent, an effective 2D dielectric permittivity $\varepsilon$ needs to be defined. Then Eqs.~(\ref{Eq:w})--(\ref{Eq:W}) can be readily extended to 2D materials, provided the free energy in Eq.~(\ref{Eq:W}) is integrated over the surface area, which eliminates ambiguities associated with the layer thickness.

\begin{table*}
    \caption{Structural parameters and effective 2D elastic, piezoelectric and static dielectric properties of single-layer hexagonal \ce{MoS2} and \ce{WS2} from self-consistent DFT calculations (relaxed-ion approximation).}\label{Table:Properties}
    \begin{ruledtabular}
        \begin{tabular}{l c c c c c}
             & & \multicolumn{2}{c}{\ce{MoS2}} & \multicolumn{2}{c}{\ce{WS2}} \\
             \cline{3-4}\cline{5-6}
            Parameter & Units & Calculated & Other sources & Calculated & Other sources \\
            \hline
            $a_0$ & {\AA} & 3.185 & 3.16\footnote{Experimental \cite{Wilson_AP_18_1969}\label{fn:1}}, 3.19\footnote{Calculated with DFT/GGA \cite{Duerloo_JPCL_3_2012}\label{fn:2}}  & 3.188 & 3.15$^\text{\ref{fn:1}}$, 3.19$^\text{\ref{fn:2}}$ \\
            $C_{11}$ & N/m & 133 & 130$^\text{\ref{fn:2}}$ & 146 & 144$^\text{\ref{fn:2}}$ \\
            $C_{12}$ & N/m & 33 & 32$^\text{\ref{fn:2}}$ & 32 & 31$^\text{\ref{fn:2}}$ \\
            $e_{11}$ & pC/m & 359 & $290\pm50$\footnote{Experimental \cite{Zhu_NN_10_2015}}, 364$^\text{\ref{fn:2}}$ & 249 & 247$^\text{\ref{fn:2}}$ \\
            $\chi_{11}$ & F & $7.4\cdot10^{-20}$ & $7.5\cdot10^{-20,}$\footnote{Obtained using Eq.~(\ref{Eq:epsilon3D}) based on \ce{MoS2} bulk in-plane relative dielectric permittivity of 15 and the interlayer separation of 6.02~{\AA} \cite{Liang_N_6_2014}.} & $7.0\times10^{-20}$ & $7.0\times10^{-20,}$\footnote{Obtained using Eq.~(\ref{Eq:epsilon3D}) based on \ce{WS2} bulk in-plane relative dielectric permittivity of 14 and the interlayer separation of 6.06~{\AA} \cite{Liang_N_6_2014}.} \\
            %$\varepsilon_{11}$ & F & $10.1\times10^{-20}$ & $(3.6-4.6)\times10^{-20}$\footnote{Obtained using the range of $6.6-8.5$ for the bulk relative dielectric permittivity and the interlayer separation of 6.15~{\AA}.} & $9.3\times10^{-20}$ & \\
        \end{tabular}
    \end{ruledtabular}
\end{table*}

Structural, elastic, piezoelectric, and dielectric properties of monolayer \ce{MoS2} and \ce{WS2} are gathered in Table~\ref{Table:Properties}. The structural unit and orientation of coordinate axes are illustrated in Fig.~\ref{Fig:Polarization}(a). The calculated lattice parameters are in agreement with experiment and other calculations reported in the literature. The hexagonal symmetry of a single layer (point group D$_\text{3h}$) reduces the number of independent coefficients in the stiffness matrix down to two: $C_{11}$ and $C_{12}$ \cite{NyeJ_1985_Physical_prop_crystals}. Our values of $C_{11}$ and $C_{12}$ listed in Table~\ref{Table:Properties} agree with those obtained in previous DFT calculations. The piezoelectric tensor is characterised by a single independent element $e_{11}$, due to symmetry arguments. The calculated values agree well with prior theoretical studies. However, approximately 20\% deviation from existing experimental data is observed. This deviation is acceptable giving the large uncertainty of experimental measurements.

The static dielectric permittivity is one of the least studied properties of single-layer transition metal dichalcogenides. The present calculations yield the value of $\varepsilon^\text{3D}_{11}/\varepsilon_0=4.5$ for the in-plane relative dielectric permittivity of a single-layer \ce{MoS2}, with $\varepsilon_0$ being the permittivity of free space. It should be emphasised that $\varepsilon^\text{3D}$ is an extensive property, which is determined by the thickness of the vacuum layer $H_\text{v}$ that is used for separation between periodical images in the direction perpendicular to the planar structure. To represent a free-standing layer of \ce{MoS2}, the value of $H_\text{v}=24.6$~{\AA} was chosen, which is approximately by a factor of four greater than the spacing between layers in bulk. Berkelbach~\textit{et al.}~\cite{Berkelbach_PRB_88_2013} proposed evaluation of the effective 2D polarizability $\chi^\text{2D}$ of planar materials using the following relationship
\begin{equation}\label{Eq:epsilon3D}
	\varepsilon^\text{3D} = \varepsilon_0 + \frac{\chi^\text{2D}}{H_\text{v}},
\end{equation}
which yields the effective in-plane polarizability of $\chi^\text{2D}_{11}=7.4\cdot10^{-20}$~F, as compared to the value of $\chi^\text{2D}_{11}=7.5\cdot10^{-20}$~F obtained for bulk \ce{MoS2} (see Table~\ref{Table:Properties}).

Potential energy profile scans similar to those shown Fig.~\ref{Fig:Energy-profile} reveal the presence of a zig-zag electric field even in the middle of the vacuum region due to periodic boundary conditions along $z$-axis (see Supplementary information). To capture the energy stored in the vacuum due to the finite electric field, the effective 2D dielectric permittivity used in calculation of the free energy density in Eq.~(\ref{Eq:w}) is expressed as
\begin{equation}\label{Eq:epsilon2D}
	\varepsilon^\text{2D} = \chi^\text{2D} + \varepsilon_0 H_\text{v}~.
\end{equation}
The additional term $\varepsilon_0 H_\text{v}$ contributes approximately 25\% to the value of $\varepsilon^\text{2D}$. 

Minimization of the total free energy $W$ for the 2D strained lateral heterostructure of \ce{MoS2} and \ce{WS2} was performed using a Lagrange multiplier approach with respect to the strain tensor $\bm{\epsilon}^\text{I,II}$ and electric field $\bm{E}^\text{I,II}$ in both domains (see Methods for details). The quasi-2D continuum model with material parameters listed in Table~\ref{Table:Properties} yields the strain distribution of $\epsilon_1^\text{I}=0.1045$ and $\epsilon_1^\text{II}=0.0955$, which is in excellent agreement with DFT results. The greater strain in \ce{MoS2} (domain~I) is due to its lower stiffness $C$ as compared to \ce{WS2} (see Table~\ref{Table:Properties}). The continuum model also properly captures magnitude of the electric field $|E|=8.2\cdot10^7$~V/m, which coincides with the average slope of the electrostatic potential profile obtained from first-principle calculations.

\begin{figure*}
    \includegraphics[width=0.95\textwidth]{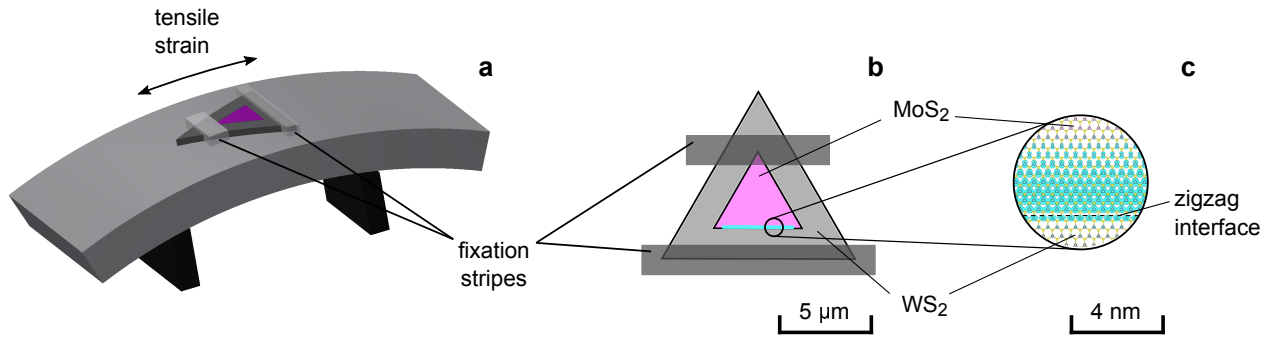}\\
	\caption{(a) Schematic illustration of a four-point bending setup for straining a triangular \ce{MoS2}/\ce{WS2} lateral heterosctructure. (b,c) 1DEG is formed at the zigzag interface oriented perpendicular to the applied strain. }\label{Fig:Test-setup}
\end{figure*}

Finally, we would like to comment on a practical realisation of the strained heterostructures discussed in this paper. \ce{MoS2}/\ce{WS2} lateral heterostructures usually have a morphology of equilateral triangular flakes of the size of a few micrometres \cite{Gong_NM_13_2014,Huang_NM_13_2014}. \ce{MoS2} forms an inner core surrounded by the \ce{WS2} outer layer \cite{Bogaert_NL_16_2016}. Gong~\textit{et~al}.\cite{Gong_NM_13_2014} reported achieving an atomically sharp \ce{MoS2}/\ce{WS2} in-plane interface. The interface is preferentially formed along ``zigzag" direction (the y-axis in Fig.~\ref{Fig:Heterostructure},a), which is consistent with the structural model studied here. The strain can be applied employing a setup shown in Fig.~\ref{Fig:Test-setup} previously used by Conley~\textit{et~al}.\cite{Conley_NL_13_2013} to measure the band gap shift of \ce{MoS2} with strain. The method involves clamping of a specimen at the surface of a mechanically bent substrate, which allows applying of a uniform strain up to 2\% in a highly controlled manner. The strain magnitude much less than 10\% can be sufficient giving a much larger length of real heterostructures in comparison to that modelled here. The presence of a strain-induced electric field can be verified by measuring a photoluminescence (PL). In unstrained \ce{MoS2}/\ce{WS2} lateral heterostructures, the PL intensity is enhanced at the \ce{MoS2}/\ce{WS2} interface \cite{Gong_NM_13_2014,Huang_NM_13_2014} due to the type-II band alignment \cite{Kang_JPCC_119_2015}. The PL intensity at the interface that develops 1DEG is expected to diminish when the strain is applied due to the induced electric field that separates charge carriers.

%-----------------------------------------------------------------------
%
%                       C O N C L U S I O N S
%
%-----------------------------------------------------------------------
\subsection*{\large Conclusions \hfill~}\vspace{-12pt}

\noindent One-dimensional conductivity channel is obtained in a lateral \ce{MoS2}/\ce{WS2} heterojunction. Conducting electronic states are confined along the interface by an inhomogeneous electric field that is induced by differences in the piezoelectric and elastic response of two materials thereby creating a one-dimensional electron gas. An effective model that captures interactions between electric and elastic degrees of freedom in low-dimensional heterostructures is developed. The model accurately predicts the magnitude of macroscopic electric field induced in the strained heterostructure as verified by \textit{ab initio} calculations. This realisation of 1D electron gas creates an alternative to a quasi 1D conducting channel formed in the 2D electron gas of GaAs/(AlGa)As heterostructures by electrostatic gating \cite{Thornton_PRL_56_1986,Berggren_PTRSA_368_2010} that can be potentially used for low-power switching applications.

%-----------------------------------------------------------------------
%
%                       A P P E N D I X 
%
%-----------------------------------------------------------------------

\subsection*{\large Methods \hfill~}\label{Sec:Method} \vspace{-12pt}

\noindent\textbf{Calculation of structural, elastic, and dielectric properties.}~~Electronic structure calculations of single-layer hexagonal \ce{MoS2} and \ce{WS2} have been performed in the framework of the density functional theory (DFT) \cite{Kohn_PR_140_1965} using Perdew-Burke-Ernzerhof generalized gradient approximation (GGA-PBE) for the exchange-correlation functional \cite{Perdew_PRL_77_1996}. Structural, elastic, and dielectric properties were modelled using the Vienna \textit{ab initio} simulation program (VASP) and projector augmented-wave (PAW) potentials \cite{Kresse_PRB_54_1996,Kresse_PRB_59_1999,Blochl_PRB_50_1994}. The structure was represented by a single layer of \ce{MoS2} or \ce{WS2} with a vacuum separation, which is approximately equal to a quadruple value of the equilibrium spacing between layers of the bulk 2H-\ce{MoS2}. The structural optimisation was carried out in conjunction with relaxation of the in-plane lattice parameter $a$. The structure was considered optimised when the magnitude of Hellmann-Feynman forces acting on atoms dropped below 2~meV/{\AA}. The Brillouin zone of the primitive unit cell was sampled using $16\times16\times1$ Monkhorst-Pack grid \cite{Monkhorst_PRB_13_1976}. The mesh was appropriately scaled when supercells are considered.

A hard PAW potential was used to represent sulphur (S\_h). Semi-core electrons were included as valence electrons in molybdenum (Mo\_sv) and tungsten (W\_pv). The cutoff energy for a plane wave expansion was set at 500~eV, which is 25\% higher than the value recommended in the pseudopotential file. The higher cutoff energy was essential for obtaining accurate, converged lattice parameters.

The elastic tensor was determined using a finite differences technique from the strain-stress relationship calculated in response to finite distortions of the lattice taking into account relaxation of the ions. The total of eight strained structures that represent various permutations of the strain $\epsilon_{1,2}=\{-0.02, 0, +0.02\}$ were considered.

The relaxed-ion dielectric tensor was calculated using the finite external electric field of the magnitude 1~meV/{\AA}. The tight energy convergence of $10^{-9}$~eV was required to achieve the accuracy of 0.1 for the relative dielectric permittivity.

\vspace{12pt}\noindent\textbf{Calculation of piezoelectric coefficients.}~~Calculations of piezoelectric coefficients were performed using a full potential linear augmented plane wave method implemented in \texttt{Wien2k} package \cite{Blaha_2001} in conjunction with \texttt{BerryPI} extension \cite{Ahmed_CPC_184_2013} that utilises a Berry phase approach \cite{King-Smith_PRB_47_1993} for computing macroscopic polarization. Piezoelectric strain coefficients are conventionally defined as
\begin{equation}\label{Eq:e_ij-definition}
    e_{ij} = \mathrm d P_i/\mathrm d \epsilon_j,
\end{equation}
where $\mathrm d P_i$ is the change in macroscopic polarisation along $i$-axes observed in response to the increment in $j$'s strain component $\mathrm d \epsilon_j$. It seems straightforward to evaluate the coefficients using a finite difference, which involves computing the polarisation of strained and unstrained structures. However, this approach introduces complications related to the choice of a reference structure that \textit{must remain commensurate} with the strained cell to serve as a reference. A similar approach was introduced by Posternak~\textit{et al.}~\cite{Posternak_PRL_64_1990} to assess the spontaneous bulk polarisation of wurtzite BeO, where the zinc blende structure served as a reference due to symmetry arguments.

In the case of hexagonal transition metal dichalcogenides, the polarisation of an unperturbed layer can be taken as a reference zero due to the cancellation of local dipoles resulted from the 3-fold rotational symmetry as illustrated in Fig.~\ref{Fig:Polarization}. Any strain tensor that preserves this symmetry (e.g., $\epsilon_1=\epsilon_2$) produces no change in polarisation. This result translates into a symmetry of the piezoelectric coefficients \cite{NyeJ_1985_Physical_prop_crystals}
\begin{equation}\label{Eq:e-symetry}
    e_{11} = -e_{12},
\end{equation}
which is inherent to D$_{3h}$ point group. It turns out that no change in the Berry phase results from the strain $\epsilon_1=\epsilon_2$. However, there is a sizeable change in polarisation originated from the increment in the cell volume that is incompatible with symmetry-imposed constraints in Eq.~(\ref{Eq:e-symetry}). To resolve this contradiction, the piezoelectric coefficients were redefined in terms of the Berry phase
\begin{equation}\label{Eq:e_ij-re-defined}
    e_{ij} = \frac{a_i}{2\pi \Omega_0}~ \frac{\mathrm d \phi_i}{\mathrm d \epsilon_j}.
\end{equation}
Here $a_i$ is the lattice parameter associated with the crystallographic axis $i$, $\Omega_0$ is the volume of the unperturbed unit cell, and $\phi_i$ is the Berry phase along direction $i$ that includes both ionic and electronic components. A least square fit technique was used to calculate piezoelectric coefficients for the total of eight strained structures that represent various permutations of the strain (the same as for elastic properties). Additional relaxation of atomic positions was performed for each stained structure.

Visualization of atomic structures was performed using \texttt{VESTA}~3 package \cite{Momma_JAC_44_2011}.

\vspace{12pt}\noindent\textbf{Free energy minimization.}~~The objective is to find a set of strains and electric fields
\begin{equation}\label{Eq:Appendix:Variables}\setcounter{MaxMatrixCols}{20}
 \begin{array}{cccccccccc}
  \epsilon_1^\text{I}, & \epsilon_2^\text{I}, & \epsilon_6^\text{I}, & 
  	E_1^\text{I}, & E_2^\text{I}, & \epsilon_1^\text{II}, & 
	\epsilon_2^\text{II}, & \epsilon_6^\text{II}, & 
	E_1^\text{II}, & E_2^\text{II}
 \end{array}
\end{equation}
that minimise the internal energy of the system $W$ defined by Eq.~(\ref{Eq:W}) for a specific case of the strained lateral heterostructure shown in Fig.~\ref{Fig:Heterostructure}.  The optimization is subject to constrains, such as an applied macroscopic strain $\epsilon_1=0.1$, continuity of both the electric displacement (Eq.~\ref{Eq:D_continuity}) and matter. From the mathematical standpoint, it is a constrained optimisation of an objective function represented by a quadratic form (Eq.~\ref{Eq:w}). The problem can be solved using a method of Lagrange multipliers.

First, a matrix is constructed to represent linear coefficients of the partial derivatives $\partial w/\partial x_k$, where $x_k$ is any variable from the list (\ref{Eq:Appendix:Variables}). When strain variables in the first domain are concerned, the linear coefficients are simply components of the elastic stiffness matrix
\begin{equation}
	\mathbb{C}^\text{I} = 
 \begin{pmatrix}
  C_{11}^\text{I} & C_{12}^\text{I} & 0 \\
  C_{12}^\text{I} & C_{11}^\text{I} & 0 \\
  0 & 0 & C_{66}^\text{I} \\
 \end{pmatrix},
\end{equation}
which is written taking into account symmetry imposed by the lattice. Similarly, the dielectric permittivity tensor 
\begin{equation}
	\mathbb{E}^\text{I} = 
 \begin{pmatrix}
  \varepsilon_{11}^\text{I} & 0 \\
  0 & \varepsilon_{11}^\text{I} \\
 \end{pmatrix}
\end{equation}
represents the linear coefficients of the partial derivatives $\partial w/\partial x_k$ for variables that correspond to the electric field components. Generalising to all optimisation variable related to the domain~I, the matrix of linear coefficients takes the form
\begin{equation}
	\mathbb{H}^\text{I} = 
	\left(
 \begin{array}{c|c}
  \mathbb{C}^\text{I} & 0 \\ 
  \hline
  0 & \mathbb{E}^\text{I}
 \end{array}
 	\right).
\end{equation}

Our objective function is not the energy density $w$, but rather the total internal energy of the system
$W$, which takes into account the individual area occupied by each domain. For the lateral junction of two domains that share the same width but may have different length $L^\text{I}$ and $L^\text{II}$ (Fig.~\ref{Fig:Heterostructure}), linear coefficients of the partial derivatives $\partial W/\partial x_k$ form a matrix
\begin{equation}
	\mathbb{H} = 
	\left(
 \begin{array}{c|c}
  L^\text{I}\mathbb{H}^\text{I} & 0 \\ 
  \hline
  0 & L^\text{II}\mathbb{H}^\text{II}
 \end{array}
 	\right).
\end{equation}

Now the following boundary conditions need to be incorporated
\begin{subequations}\label{Eq:Appendix:Boundary-conds}
\begin{align}
	e_{11}^\text{I}\epsilon_1^\text{I} + 
    	e_{12}^\text{I}\epsilon_2^\text{I} + 
		\varepsilon_{11}^\text{I}E_{1}^\text{I} - & \nonumber \\
		e_{11}^\text{II}\epsilon_1^\text{II} - 
		e_{12}^\text{II}\epsilon_2^\text{II} -
		\varepsilon_{11}^\text{II}E_{1}^\text{II} & = 0, \\
	\epsilon_1^\text{I}L^\text{I} + \epsilon_1^\text{II}L^\text{II} & = 
    	\epsilon_1(L^\text{I}+L^\text{II}), \\
	a_0^\text{I}\epsilon_2^\text{I} - a_0^\text{II}\epsilon_2^\text{II} & = 
		a_0^\text{II} - a_0^\text{I}.
\end{align}
\end{subequations}
The first condition stems from Eqs.~(\ref{Eq:D_l}) and (\ref{Eq:D_continuity}) that capture the essence of piezoelectric coupling between the strain and electric field. It is implied that the spontaneous polarisation is zero in both materials $(P_0=0)$ when unstrained. The second and third requirements account for the continuity of the heterostructure along the direction of the applied strain and perpendicular to that. The difference $a_0^\text{II} - a_0^\text{I}$ corresponds to a lattice mismatch between two materials. The left-hand-side of Eq.~(\ref{Eq:Appendix:Boundary-conds}) can be transformed into a matrix form
\begin{equation}
	\mathbb{B} = 
 \begin{pmatrix}
  e_{11}^\text{I} & -e_{11}^\text{I} & 0 & \varepsilon_{11}^\text{I} & 0 & -e_{11}^\text{II} & e_{11}^\text{II} & 0 & -\varepsilon_{11}^\text{II} & 0 \\
  L^\text{I} & 0 & 0 & 0 & 0 & L^\text{II} & 0 & 0 & 0 & 0 \\
  0 & a_0^\text{I} & 0 & 0 & 0 & 0 & -a_0^\text{II} & 0 & 0 & 0
 \end{pmatrix},
\end{equation}
where columns correspond to the optimisation variables in Eq.~(\ref{Eq:Appendix:Variables}). The symmetry of piezoelectric strain coefficients $(e_{11} = -e_{12})$ is taken into account during this transformation.

Finally, the energy terms and constraints are combined in a matrix 
\begin{equation}
	\mathbb{L} = 
	\left(
 \begin{array}{c|c}
  \mathbb{H} & \mathbb{B}^\text{T} \\ 
  \hline
  \mathbb{B} & 0
 \end{array}
 	\right)
\end{equation}
that represents Lagrangian of the problem $\mathcal{L}$. Unknowns
\begin{eqnarray}\setcounter{MaxMatrixCols}{20}
	\mathbb{X}^\text{T} & = &
 	\left(
  		\epsilon_1^\text{I},~\epsilon_2^\text{I},~\epsilon_6^\text{I},~
  		E_1^\text{I},~E_2^\text{I},~\epsilon_1^\text{II},~
		\epsilon_2^\text{II},~\epsilon_6^\text{II},~
		E_1^\text{II},~E_2^\text{II},
	\right. \nonumber \\
 & &
 	\left.
		\lambda_1,~\lambda_2,~\lambda_3
	\right)
\end{eqnarray}
are obtained by solving a linear equation
\begin{equation}
	\mathbb{L} \cdot \mathbb{X} = \mathbb{R}
\end{equation}
with the right hand side being a column vector
\begin{eqnarray}\setcounter{MaxMatrixCols}{20}
	\mathbb{R}^\text{T} & = &
		\left(
 			0,~0,~0,~0,~0,~0,~0,~0,~0,~0,~0, 
		\right.\nonumber \\
		& & 
		\left.
			\epsilon_1 (L^\text{I}+L^\text{II}),~a_0^\text{II}-a_0^\text{I}
		\right).
\end{eqnarray}
The first ten elements of $\mathbb{R}$ are zero due to the requirement of $\partial\mathcal{L}/\partial x_k=0$ at the optimum for each variable listed in Eq.~(\ref{Eq:Appendix:Variables}). The remaining elements represent the right hand side of Eq.~(\ref{Eq:Appendix:Boundary-conds}). Here $\lambda$'s are Lagrange multipliers.

\vspace{12pt}\noindent\textbf{Data availability.}~~Crystallographic information files (CIF) with atomic structures used in calculations can be accessed through the Cambridge crystallographic data centre (CCDC deposition numbers 1520213--1520216). 

%                       B I B L I O G R A P H Y
%
%-----------------------------------------------------------------------
%\clearpage
%\bibliographystyle{naturemag}
%\bibliography{../bibliography}

%-----------------------------------------------------------------------
%
%                       A C K N O W L E D G E M E N T
%
%-----------------------------------------------------------------------
\subsection*{\large Acknowledgments \hfill~}\vspace{-12pt}
\noindent Funding was provided by the Natural Sciences and Engineering Research Council of Canada under the Discovery Grant Program RGPIN-2015-04518. The work was performed using computational resources of the Thunder Bay Regional Research Institute, Lakehead University, and Compute Canada (Calcul Quebec).

\subsection*{\large Additional information \hfill~}\vspace{-12pt}
\noindent Supplementary information is available in the online version.

\subsection*{\large Competing financial interests \hfill~}\vspace{-12pt}
\noindent The author declares no competing financial interests.

\end{document}